\documentclass[journal]{aa}

\usepackage[dvips]{graphicx}
\usepackage{amsmath}
\usepackage{amssymb}
\usepackage{verbatim}
\usepackage{array}
\usepackage{txfonts}
\usepackage{natbib}

\bibpunct{(}{)}{;}{a}{}{,}

\begin{document}

% Title, header, abstract
\title{Constraints on the kinematics of the $^{44}$Ti ejecta\\ of Cassiopeia A from INTEGRAL/SPI}
\titlerunning{Constraints on the kinematics of the $^{44}$Ti ejecta of Cassiopeia A from INTEGRAL/SPI}
\author{Pierrick Martin \inst{1,2} \and J\"urgen Kn\"odlseder \inst{1,2} \and Jacco Vink \inst{3} \and Anne Decourchelle \inst{4} \and Matthieu Renaud \inst{5}}
\institute{Centre d'Etude Spatiale des Rayonnements (CESR), 9, avenue colonel Roche, BP44346, 31028 Toulouse cedex 4, France
              \and Universit\'e de Toulouse (UPS), Centre National de la Recherche Scientifique (CNRS), UMR 5187
              \and Astronomical Institute, University Utrecht, PO Box 80000, 3508 TA Utrecht, The Netherlands
              \and Laboratoire AIM, CEA/DSM-CNRS-Universit\'e Paris Diderot, DAPNIA/Sap, CEA-Saclay, 91191 Gif-sur-Yvette, France
              \and Laboratoire APC, CNRS-UMR 7164-Universit\'e Paris Diderot, 10, rue A. Domon et L. Duquet, 75205 Paris Cedex 13, France}           
\date{Received 7 March 2008 / Accepted 3 April 2009}
\abstract{The medium-lived $^{44}$Ti isotope is synthesised by explosive Si-burning in core-collapse supernovae. It is extremely sensitive to the dynamics of the explosion and therefore can be used to indirectly probe the explosion scenario. The young supernova remnant Cassiopeia A is to date the only source of gamma-ray lines from $^{44}$Ti decay. The emission flux has been measured by CGRO/COMPTEL, BeppoSAX/PDS and INTEGRAL/IBIS.}{The high-resolution spectrometer SPI on-board the INTEGRAL satellite can provide spectrometric information about the emission. The line profiles reflect the kinematics of the $^{44}$Ti in Cassiopeia A and can thus place constraints on its nucleosynthesis and potentially on the associated explosion process.}{Using 4 years of data from INTEGRAL/SPI, we have searched for the gamma-ray signatures from the decay of the $^{44}$Ti isotope. The overwhelming instrumental background noise required an accurate modelling and a solid assessment of the systematic errors in the analysis.}{Due to the strong variability of the instrumental background noise, it has not been possible to extract the two lines at 67.9 and 78.4\,keV. Regarding the high-energy line at 1157.0\,keV, no significant signal is seen in the 1140-1170\,keV band, thereby suggesting that the line signal from Cassiopeia A is broadened by the Doppler effect. From our spectrum, we derive a $\sim$ 500\,km\,s$^{-1}$ lower limit at 2$\sigma$ on the expansion velocity of the $^{44}$Ti ejecta.}{Our result does not allow us to constrain the location of $^{44}$Ti since the velocities involved throughout the remnant, derived from optical and X-ray studies, are all far above our lower limit.}

\maketitle

% Introduction
\section{Introduction}
\label{introduction}

Until very recently, Cassiopeia A (hereafter Cas A) occupied a specific place among galactic supernova remnants as being the result of the most recent galactic supernova known to date \citep[the youngest one might now be G1.9+0.3, with an estimated age of about a century, see][]{Reynolds:2008}. Based on an optical study of outer high-velocity knots, \citet{Thorstensen:2001} derived an explosion date of 1671 or slightly later for Cas A. Combined with the 3.4\,kpc distance estimated by \citet{Reed:1995}, the explosion that eventually lead to Cas A might well be the transient 6th magnitude star noticed by Flamsteed in 1680. From the light-echo scattered by interstellar dust, the event has now been identified as an SNIIb, which indicates that the progenitor was a red supergiant that had lost most of its hydrogen envelope at the moment of collapse \citep{Krause:2008}.\\
\indent Cas A retains interesting features that could be valuable clues in the long-standing effort to understand the supernova explosion mechanism. Indeed, growing evidence has arisen that the explosion and/or the expansion of Cas A were asymmetric and highly turbulent. X-ray images taken with Chandra have revealed a jet (and to a lesser extent a counter-jet) protruding in the north-east direction far beyond the nearly spherical forward shock \citep{Vink:2004,Hwang:2004}. Another indication for an asymmetric process is the substantial mixing suggested by the Chandra X-ray images in which Fe-rich material appears, in at least two extended regions, at a greater projected radius than Si-rich material \citep{Hughes:2000,Hwang:2000}. On a smaller scale, the wide variety in the so-called optical Fast-Moving Knots (FMKs) spectra regarding the relative strengths of O and S lines, together with the absence of a correlation between these spectral properties and the velocity of the FMKs, also imply some turbulence or uneven burning \citep{Chevalier:1979}.\\
\indent In addition to these peculiarities, Cas A is known as the only source of gamma-ray line emission from $^{44}$Ti decay. The unstable $^{44}$Ti isotope is a by-product of explosive Si-burning resulting from alpha-rich freeze-out \citep{Woosley:1973,The:1998}. The $^{44}$Ti is expected to lie deep in the ejecta and its yield is very sensitive to the early dynamics of the explosion and to the position of the mass cut (the boundary that separates the ejected material from the material that falls back onto the compact object). As such, $^{44}$Ti provides indirect evidence about the flow pattern that mediated the explosion energy and can therefore be used to probe the explosion scenario.\\
\indent After a mean lifetime of 85 yrs \citep{Ahmad:2006}, $^{44}$Ti decays by electron capture to $^{44}$Sc and then to $^{44}$Ca, thereby emitting three de-excitation photons at 67.9, 78.4 and 1157.0 keV. Given its youth and relative proximity, Cas A constitutes an ideal target to search for the characteristic decay emission of the medium-lived $^{44}$Ti. The first detection of the $^{44}$Ti emission from Cas A was reported by \citet{Iyudin:1994} from CGRO/COMPTEL observations at 1157.0\,keV, with a downward revision of the flux by \citet{Dupraz:1997}. Subsequent detections of the low-energy lines by BeppoSax/PDS \citep{Vink:2001} and recently by INTEGRAL/IBIS/ISGRI \citep{Renaud:2006} have allowed investigators to tighten the flux value for the $^{44}$Ti emission from Cas A to (2.5 $\pm$ 0.3) $\times 10^{-5}$ ph\,cm$^{-2}$\,s$^{-1}$ \citep{Renaud:2006}. With the age and distance mentioned above for Cas A and assuming that the $^{44}$Ti is not strongly ionized \citep[which could delay its decay by electron capture, see][]{Mochizuki:1999}, this flux translates into an initial $^{44}$Ti mass of about 1.6 $\times 10^{-4}$ M$_\odot$.\\
\indent While the total $^{44}$Ti flux, and hence the total $^{44}$Ti mass, is rather well constrained today, little is known about the kinematics of the $^{44}$Ti ejecta as none of the above-mentioned instruments had the capabilities required for fine spectrometry. The high-resolution spectrometer SPI on-board the INTEGRAL satellite opened the way for a determination of line profiles and thus for an estimate of the $^{44}$Ti velocity. In particular, the high-energy line at 1157.0\,keV is expected to reflect the motion of the $^{44}$Ti ejecta. Such a measurement is essential to achieve a complete understanding of the nucleosynthesis of $^{44}$Ti in Cas A and is relevant to fundamental questions such as how core-collapse supernovae explode and why Cas A is the only source of $^{44}$Ti emission in the sky \citep{The:2006}.\\
\indent In this paper we present the INTEGRAL/SPI observations of the $^{44}$Ti decay line at 1157.0\,keV and their implications in terms of velocity of the $^{44}$Ti ejecta. From this constraint, the site of the nucleosynthesis of $^{44}$Ti is then discussed in the context of all other observational data available on Cas A.

% Data analysis
\section{Data analysis}
\label{analysis}

% Instrument
\subsection{Instrument characteristics}
\label{analysis_instru}

The SPI instrument onboard the INTEGRAL gamma-ray space observatory is a high-resolution spectrometer with imaging capabilities \citep{Vedrenne:2003,Roques:2003}. The spectrometry of gamma radiation with energies between 20 keV and 8 MeV is performed by an array of 19 high-purity germanium detectors, with a spectral resolution of about 2 keV FWHM at 1 MeV. All gamma-ray sources, diffuse or point-like, emitting in this energy range can also be imaged indirectly thanks to a coded mask system with an angular resolution of 2.8$^{\circ}$ and a fully-coded field of view of 16$^{\circ}$x16$^{\circ}$. For a given pointing, a coded-mask achieves a spatial modulation of the celestial signal and the reconstruction of the sky intensity distribution exploits the correlations between the recorded image and the mask pattern or a decoding array derived from it \citep[see][]{Fenimore:1978}. The instrument response of SPI, however, is quite complex and inversion by correlation methods is not possible. More information should be accumulated and the observation of an astrophysical object or region with SPI therefore consists of a series of pointings around the target, following a so-called Òdithering patternÓ (thus adding a temporal modulation of the astrophysical signal). The sky signal is then reconstructed through a model-fitting approach.\\ 
\indent The photon interactions recorded by SPI are divided into two classes: single events, be they from astrophysical or instrumental origin, depositing all of their energy in a single detector, and multiple events, in which case the interactions involve several adjacent detectors (which therefore form a so-called pseudo-detector). In the work presented here, only single events (hereafter SE) and double events (hereafter ME2) were considered. Detector 2 and 17 failed during orbit 142 and 210 respectively. Apart from a reduction in effective area, this led to an increase of the background in the SE recorded by the detectors adjacent to the failed ones.

%Data set
\subsection{Data set}
\label{analysis_data}

\begin{figure*}[p]
\centering
\includegraphics[width=15cm]{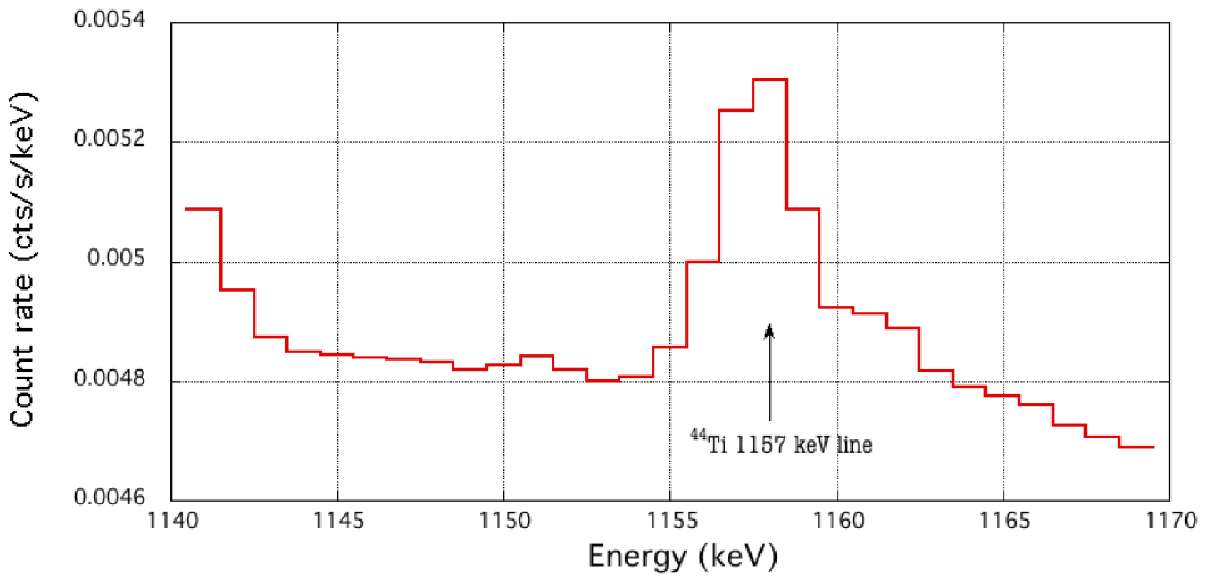}
\includegraphics[width=15cm]{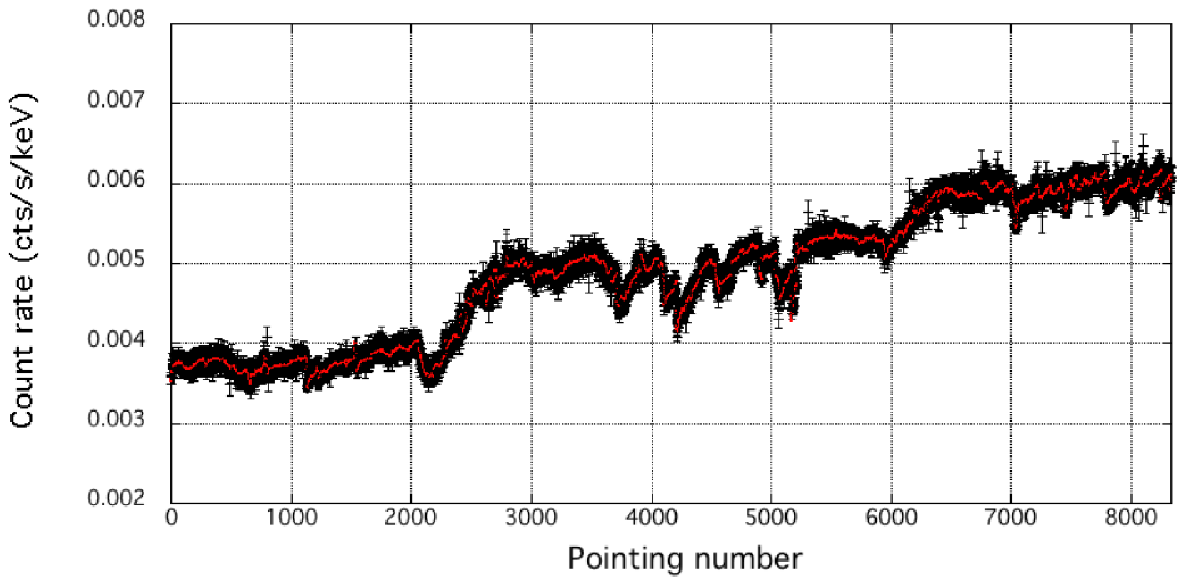}
\includegraphics[width=15cm]{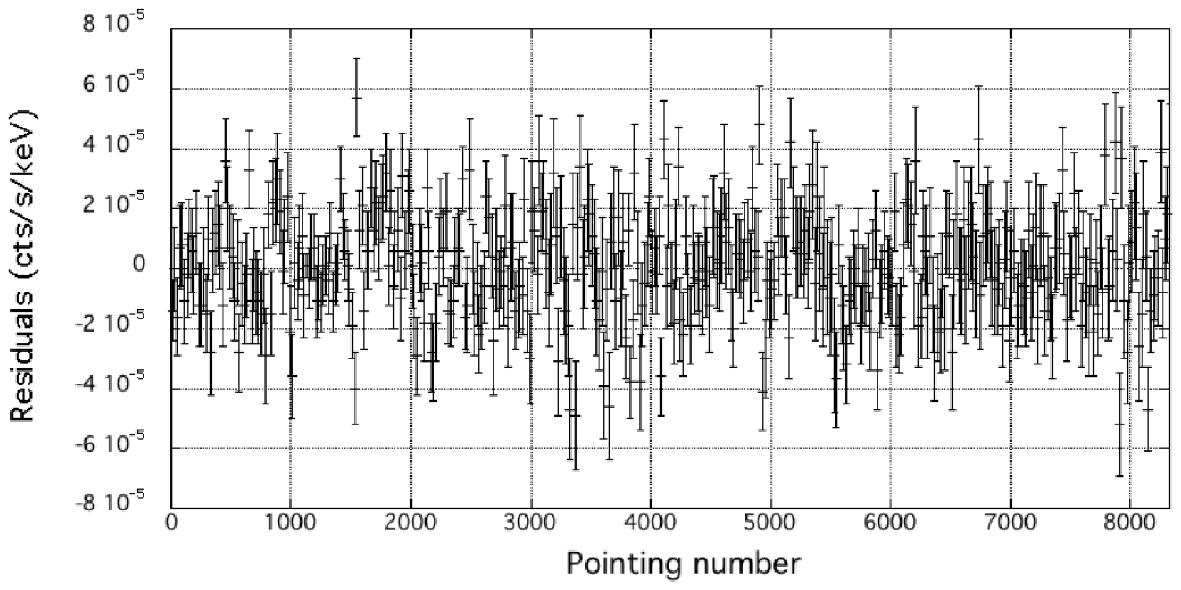}
\caption{Raw SPI spectrum with arrow indicating the nominal energy of the high-energy $^{44}$Ti line (top). Mean count rate in the 1140-1170 keV band as a function of pointing: black crosses are SPI data and the red line is the sum of background and source models (middle). Residual count rate as a function of pointing after subtraction of the fitted background and source models to the data, pointings have been rebinned in groups of 20 for clarity (bottom).}
\label{rawdatagamma}
\end{figure*}

An observation group (hereafter OG) has been built from all pointings within 20$^{\circ}$ around Cas A achieved between revolution 7 and revolution 539, plus all pointings outside the galactic plane ($\mid$b$\mid$ $\geq$ 25$^{\circ}$) during the same period. The former are mainly dedicated observations of the Cas A remnant complemented by some pointings from the periodic Galactic Plane Scan, while the latter are empty fields (at the energy considered) that will help to constrain the background level. All SE and ME2 data with energies between 1130 and 1200 keV were considered (ME2 account for about a quarter of all counts at these energies). This initial selection of pointings has been screened for abnormally high counting rates such as those occurring as the satellite crosses the radiation belts or experiences solar flares. The eventual OG is a three-dimensional space with the following characteristics:  8328 pointings (time axis), 19 single detectors and 42 double detectors (space axis) and 70 one-keV bins ranging from 1130 to 1200 keV (energy axis). This amounts to 6.65 Ms of effective observation time towards Cas A and 12.5 Ms of effective observation time of empty regions, spread over a 4.2 year interval between December 2002 and March 2007.\\
\indent The corresponding raw spectrum and count rate are presented in Fig. \ref{rawdatagamma}. Most of these reflect the instrumental background noise, the origin and handling of which are detailed in the following section.

% Background
\subsection{Instrumental background noise}
\label{analysis_bgm}

The constant bombardment of the satellite by solar and cosmic-ray particles generates a very strong background noise through the constant activation or excitation of the on-board material. This background noise manifests in the form of fluorescence and nuclear lines on top of a continuum \citep{Jean:2003}. It overwhelms the astrophysical signal by a factor of 100 and exhibits time, spectral and even spatial variations within the detector plane. The strong time variability of the background noise is illustrated in the middle plot of Fig. \ref{rawdatagamma}, where the increasing trend is due to the evolution of the solar activity that modifies the cosmic-ray flux at the satellite position, while the discontinuities are the result of the selected pointings of the OG being non-contiguous in time. In addition, some of the instrumental background lines correspond exactly to the transitions searched for by the nuclear gamma-ray astronomers. In particular, a background line exists at 1157.0 keV (see upper plot of Fig. \ref{rawdatagamma}), and corresponds to the production and subsequent nuclear de-excitation of $^{44}$Ca in the instrument. Care should therefore be taken when analysing the data to ensure complete background line subtraction.\\
\indent Standard methods for analysing the data of a coded-mask instrument make use of \textit{detector patterns}, which give the relative background count rates in each detector at a given energy and are calibrated from empty field observations. For a given observation, the \textit{detector pattern} is scaled to the data and the excess counts are then accounted for by fitting a source model or exploited by some image reconstruction algorithm. In the specific case of astrophysical gamma-ray lines, however, the signals are so weak that this approach becomes prohibitive. The variations of the background noise level from pointing to pointing are of the order of the searched signals and the relative count rates in the detectors are subject to some long-term evolution. The search for a weak source using a \textit{detector pattern} would therefore require a frequent scaling of the pattern (for each pointing in fact) and a few readjustments of it to cope with long-term variations. With such a large number of degrees of freedom, the resulting sensitivity is far too low for most astrophysical gamma-ray lines. Bringing the signals to the fore thus necessitates alternative methods involving fewer parameters so as to improve the sensitivity of the instrument. The true challenge here is the reproduction of the time evolution of the background noise, at each energy and for each detector.\\
\indent In the approach we have used here, the background is modelled as a linear combination of various \textit{templates}, which are time-series intended to account for the different sources and temporal trends of the background noise. It should be emphasized, however, that despite all efforts \citep[see for example][]{Weidenspointner:2003}, the true nature of the background remains mostly unknown, and as a consequence our approach is empirical. The central element of this stage is the so-called GEDSAT activity tracer, which represents the count rates of all events that saturate the detectors, that is all events with incoming energies above 8 MeV. It gives an estimate of the flux of high-energy cosmic-ray (and solar) particles, and therefore traces the instant activation or excitation rate of the on-board material. The use of the GEDSAT in our modelling is an hypothesis on the time variability of the real background and, as such, it could introduce some bias in the results. Yet, many analyses of SPI gamma-ray line observations have been performed under this prescription \citep[see][for the annihilation, $^{60}$Fe and $^{26}$Al lines respectively]{Knodlseder:2005,Wang:2007,Diehl:2006} and produced successful results, so that adopting a similar strategy for the study of the $^{44}$Ti lines seems justified; in addition we will ensure that the results are not affected by systematic effects through a thorough analysis of the residuals.\\
\indent For the low-energy lines at 67.9 and 78.4\,keV, the background noise turned out to be so high and so variable that it was impossible to build a satisfactory model. For the analysis of the 1157.0\,keV line, on the contrary, the background proved to be quite well represented by a simple three-component background model of the following form:
\begin{equation}
\begin{split}
BGM_{p,d,e} = & \quad A_{j(p,d,e)} \times LIVE_{p,d}\\
& + B_{k(p,d,e)} \times (T_{p}-T_{0}) \times LIVE_{p,d}\\
& + C_{l(p,d,e)} \times GEDSAT_{p,d} \times LIVE_{p,d}
\end{split}
\end{equation}
In the above formula, $BGM$ is the resulting background model and the indices $p$, $d$ and $e$ indicate the pointing, the detector and the energy bin or band considered. $A$, $B$ and $C$ are the scaling coefficients of each component, and the functions $j$, $k$ and $l$ define the parameter set for each component. $T_p$ and $T_0$ are respectively the mean date of pointing $p$ and the date of the beginning of the mission, and $GEDSAT_{p,d}$ and $LIVE_{p,d}$ are the rate of saturating events and the livetime of detector $d$ during pointing $p$. The first component (controlled by factors $A_j$) is a constant component, supposedly accounting for the part of the background that originates from the long-lived natural radioactivity of the elements making up the satellite. The second component (controlled by factors $B_k$) is a linear growth from the day the satellite was put into space, and accounts for any build-up of a specific medium-lived radioisotope due to a continuous activation by cosmic-ray bombardment. The third component (controlled by factors $C_l$) models the prompt component of the cosmic-ray induced background, that is, all background radiation arising from short-lived nuclei or excitation states (short compared to the typical duration of a pointing). The $A_j$, $B_k$ and $C_l$ factors cannot be estimated \textit{a priori} with sufficient accuracy and should therefore be determined through fitting to the data.

% Model fitting
\subsection{Model fitting}
\label{analysis_fit}

\begin{figure}[t!]
\centering
\includegraphics[width=9cm]{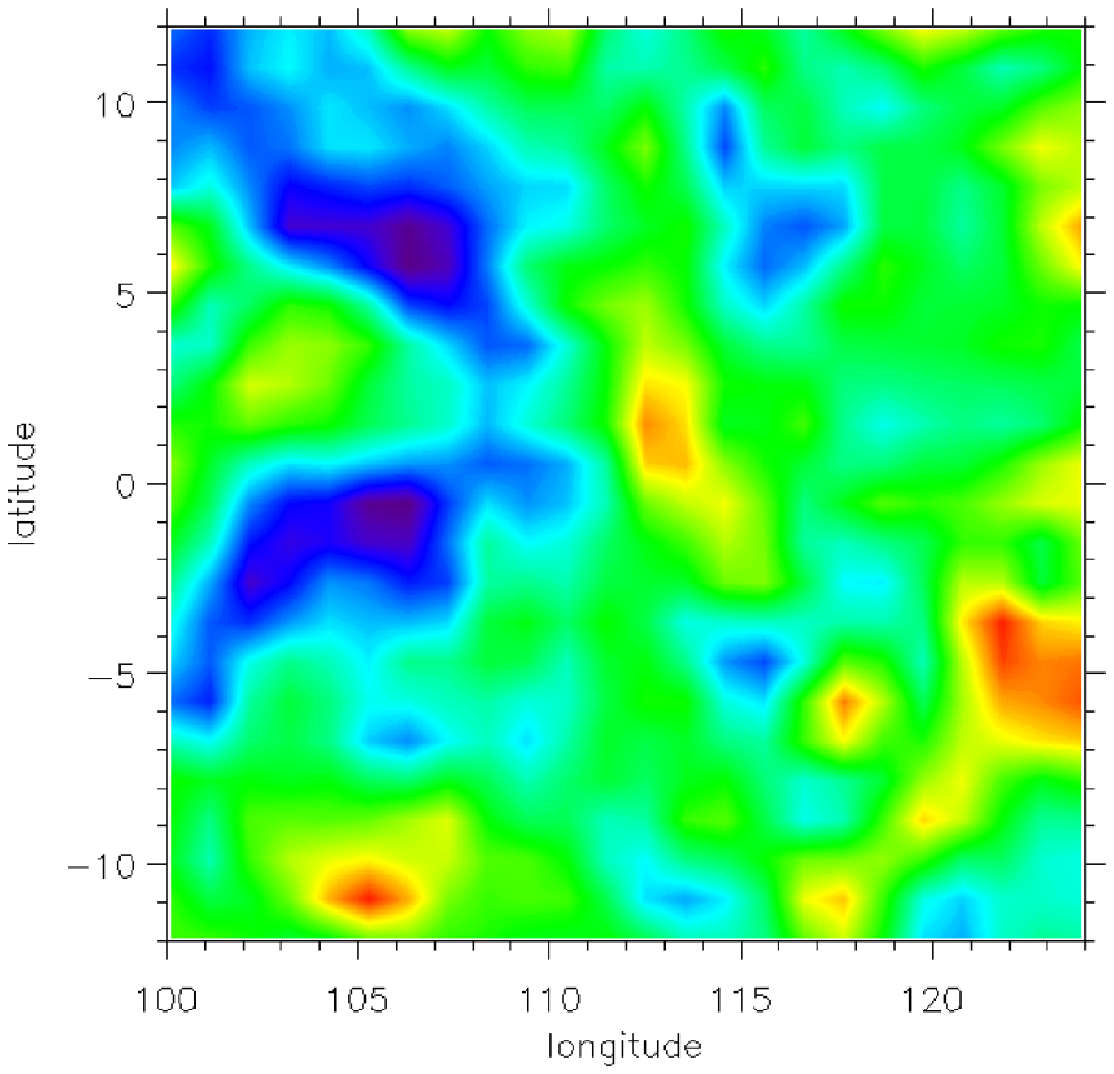}
\includegraphics[width=9cm]{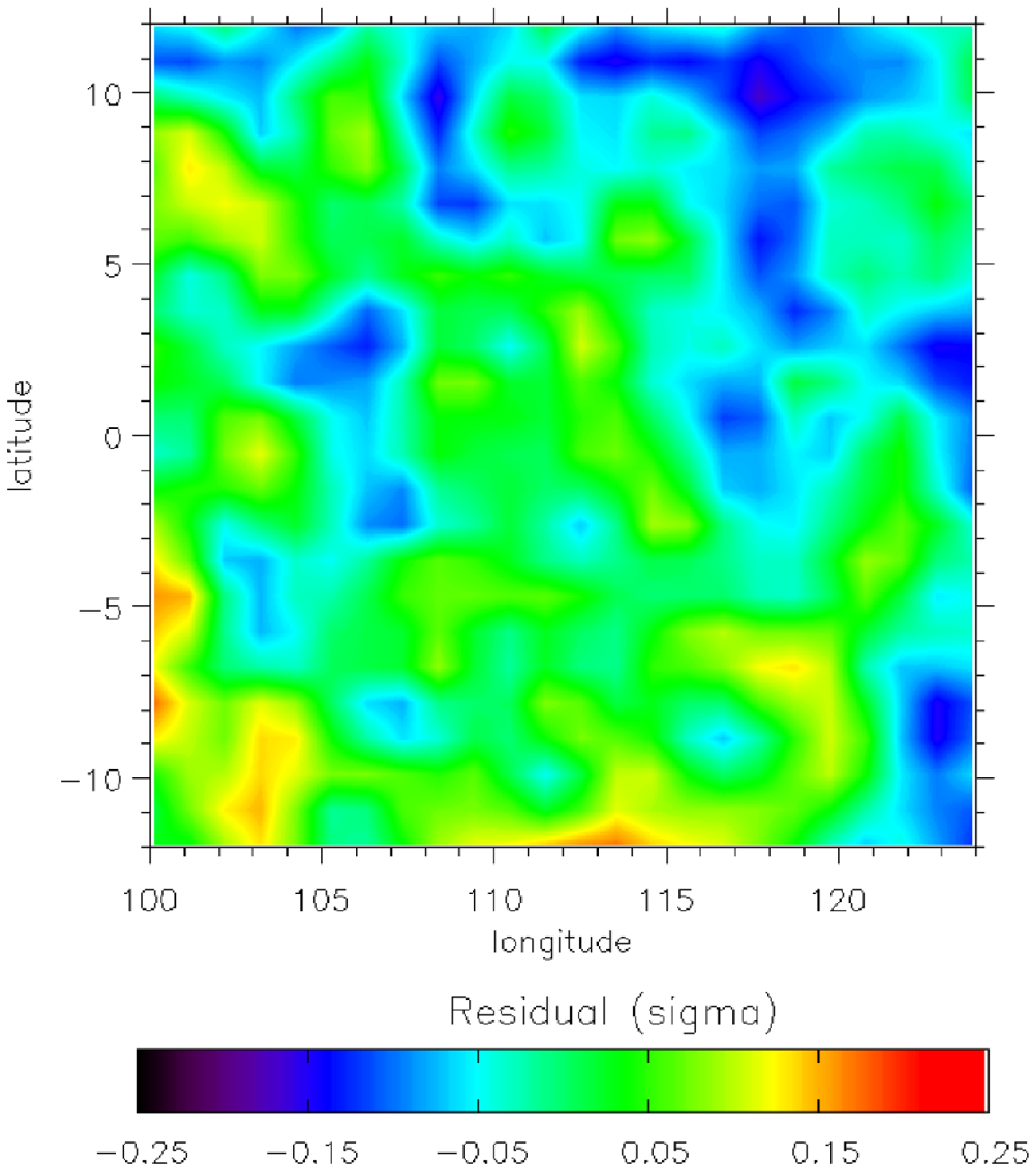}
\caption{Images of the sky residuals from our analysis (top) and from a typical simulation (bottom).}
\label{skyres_image}
\end{figure}
In this work, we want to obtain spectral information on a fiducial source of $^{44}$Ti emission whose position is well established. The $^{44}$Ti signal is therefore extracted through model fitting (as opposed to image reconstruction). Cas A is considered as a point source at (l,b)=(111.73$^{\circ}$,-2.13$^{\circ}$), since the size of the remnant \citep[roughly 5' on the Chandra images from][]{Hwang:2004} is far below the SPI angular resolution. This source model is mapped into the data-space through a convolution with the instrument response function and then forms the sky model. All background and sky components are then fitted simultaneously to the data using a Maximum Likelihood criterion for Poisson statistics. The source model is fitted independently for each energy bin to extract the source spectra, while each component of the background model is fitted independently for each detector, energy bin and for 3 time periods delimited by the failures of detectors 2 and 17 (because the failures have changed the background noise in the detectors surrounding the dead ones qualitatively and quantitatively).

% Systematics
\subsection{Assessment of systematic effects}
\label{systematics}

\begin{figure}[t]
\centering
\includegraphics[width=9cm]{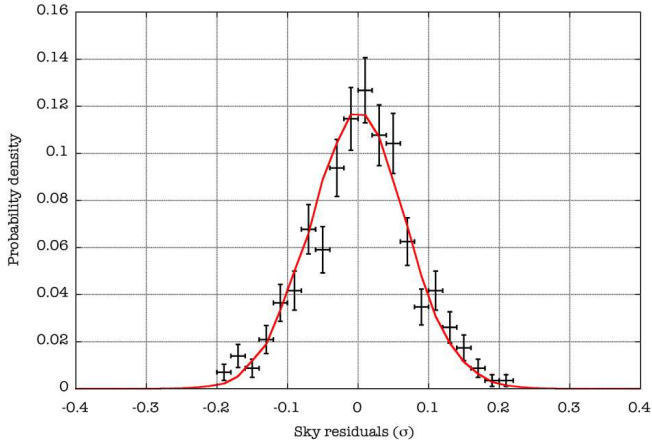}
\caption{Distribution of the sky residuals obtained from our analysis (black crosses, only non-zero points are shown) compared to the expected statistical distribution obtained from simulations (red curve).}
\label{skyres_distrib}
\end{figure}
Before proceeding with the interpretation of the result, it is important to check that our analysis of the SPI data is satisfactory. We want to be sure that our modelling of the background has not introduced systematic errors that might have prevented the detection of a signal.\\
\indent The residuals left from the subtraction of all fitted models (background and sky) to the data are projected along the time and energy dimensions of the data-space. It is checked using chi-square tests that the residuals are statistically distributed and do not suspiciously group over certain periods or energy ranges. The temporal residuals for the present analysis are shown in the lower plot of Fig. \ref{rawdatagamma}, where they have been rebinned for clarity. The reduced $\chi^2$ when residuals are determined on a pointing basis is 1.0. Yet, when computing the residuals over increasing time-scales, the reduced $\chi^2$ increases up to values of about 2.0, indicating that long-term systematic effects are present. Some long-term trends may not be properly accounted for by our background model, which can be explained by the fact that our main ingredient, the GEDSAT rate, traces instant activity only and not build-up effects. However, the propagation of these systematics from the data-space to the final results is not straightforward. Firstly, the instrument response is quite complex and image reconstruction is based on specific correlations between detectors. Secondly, when working with large OG including several thousand pointings and spanning a long time range, it may well be that systematic errors average out.\\
\indent In order to assess the impact of the systematic errors on the result, we have studied the distribution of sky residuals. Sky residuals correspond to a back-projection of the residuals from the detector plane to the sky, through the coded mask. It is the sky space, rather than the data space, that we are interested in, so that this analysis tells us more than the temporal residuals discussed previously. Formally, sky residuals result from the following calculation:
\begin{align}
\label{eq_skyres}
I_{\alpha,\delta} &=\frac{(N_{\alpha,\delta}-U_{\alpha,\delta})}{\sqrt{U_{\alpha,\delta}}} \\
\textrm{ with   } N_{\alpha,\delta} &=N_{norm}*\sum_{p,d} R_{\alpha,\delta,p,d} \times n_{p,d} \notag \\
\textrm{ and   } U_{\alpha,\delta} &=U_{norm}*\sum_{p,d} R_{\alpha,\delta,p,d} \times \mu_{p,d} \notag \\
\textrm{ where  } N_{norm} &= \frac{\sum_{p,d} n_{p,d}}{\sum_{\alpha,\delta} \sum_{p,d} R_{\alpha,\delta,p,d} \times n_{p,d}} \notag \\
\textrm{ and   } U_{norm} &= \frac{\sum_{p,d} \mu_{p,d}}{\sum_{\alpha,\delta} \sum_{p,d} R_{\alpha,\delta,p,d} \times \mu_{p,d}} \notag
\end{align}
where $I_{\alpha,\delta}$ is the image of sky residuals, expressed in number of $\sigma$. $N_{\alpha,\delta}$ and $U_{\alpha,\delta}$ are respectively the data counts and model (background and sky) counts back-projected on the plane of the sky from the $n_{p,d}$ and $\mu_{p,d}$, which are respectively the data and model counts recorded by detector $d$ during pointing $p$, for a given energy range. $R_{\alpha,\delta,p,d}$ is the instrument response function (IRF) that gives for each direction in the sky the expected distribution of events over the detector plane (derived from \citet{Sturner:2003}). A normalisation is necessary to ensure that the number of back-projected counts is identical to the number of counts in the data space.\\
\indent  To allow qualitative and quantitative assessment of our sky residuals, simulations were performed for the same OG, using as the true background function our fitted background model and as the true source function a point-source at the position of Cas A with a flux of 2.5 $\times 10^{-5}$ ph\,cm$^{-2}$\,s$^{-1}$. The sky residuals obtained in the 1150-1165\,keV band are presented in Fig. \ref{skyres_image}, together with a typical simulation. The images are qualitatively similar, suggesting that our residuals are close to what is expected from statistical fluctuations only. This is better illustrated in Fig. \ref{skyres_distrib}, where the distribution of our residuals is compared to the purely statistical distribution obtained from 10 simulations. Both distributions match, thereby indicating that the systematic errors revealed by the temporal residuals do not convert into any bias on the sky properties. We can therefore consider that our $\sim$ 500\,km\,s$^{-1}$ lower limit on the expansion velocity of the $^{44}$Ti ejecta is robust.

% Results and discussion
\section{Results and discussion}
\label{discussion}

% The spectrum of Cassiopeia A
\subsection{The spectrum of Cassiopeia A}
\label{discu_result}

\begin{figure}[t]
\centering
\includegraphics[width=9cm]{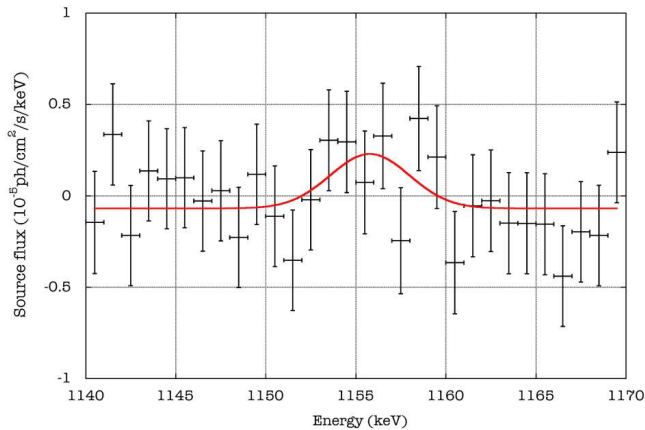}
\caption{Cassiopeia A spectrum at 1157.0\,keV combining SE and ME2; the red solid curve is the fit of a Gaussian shape.}
\label{spectra}
\end{figure}
\indent The model-fitting procedure described in \ref{analysis_fit} yields the source spectrum presented in Fig. \ref{spectra}, where SE and ME2 have been combined. The spectrum has been fitted by a Gaussian shape with a flat continuum underneath and the best-fit profile corresponds to a line flux of (1.6 $\pm$ 1.2) $\times 10^{-5}$ ph\,cm$^{-2}$\,s$^{-1}$ at a position of 1155.8 $\pm$ 1.5\,keV. This is definitely not significant (although consistent with the expected flux and position within the error bars) and it suggests that the 1157.0\,keV line signal from Cas A may be broadened by the Doppler effect, with its flux spread over a too wide energy range and consequently lost in the statistical fluctuations. From this nearly flat spectrum we can derive a statistical lower limit on the line width.\\
\begin{figure}[t]
\centering
\includegraphics[width=9cm]{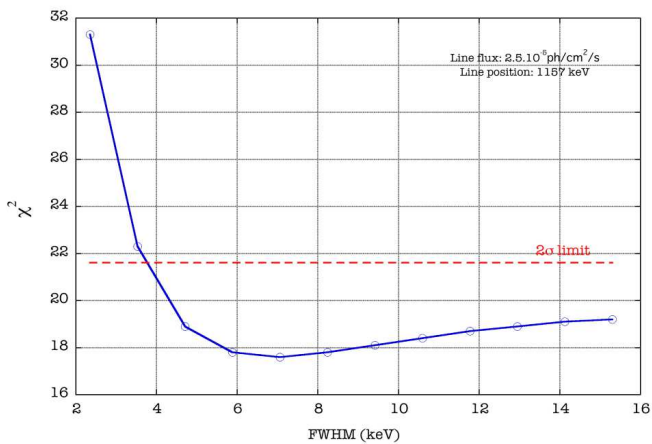}
\caption{$\chi^2$-curve for the total line width (including intrinsic and instrumental broadening), assuming a line flux of 2.5 $\times 10^{-5}$ ph\,cm$^{-2}$\,s$^{-1}$ and no position shift.}
\label{khi2curve}
\end{figure}
\indent Assuming a total line flux of 2.5 $\times 10^{-5}$ ph\,cm$^{-2}$\,s$^{-1}$ and no position shift, the 2$\sigma$ lower limit on the measured line width is 3.8\,keV FWHM (see the $\chi^2$-curve shown in Fig. \ref{khi2curve}). The spectral shape eventually appearing in the source spectrum results from the convolution of the intrinsic astrophysical 1157.0\,keV line profile with the spectral response of the instrument. Under the assumption that both profiles can be described by Gaussian functions, the measured line width is a quadratic sum of the instrumental and astrophysical line widths. Since the spectral resolution of SPI at 1\,MeV is about 2\,keV FWHM, our lower limit on the measured line width translates into a lower limit on the astrophysical width of about 3.2\,keV FWHM, which in turn implies an $\sim$500\,km\,s$^{-1}$ Doppler velocity at 1157.0\,keV. This value is an order of magnitude lower limit on the expansion velocity of the $^{44}$Ti ejecta, since our spectrum clearly does not allow us to discriminate between different sophisticated velocity patterns.

% The iron and velocity distribution in Cassiopeia A
\subsection{The iron and velocity distribution in Cassiopeia A}
\label{discu_iron}

\indent From SPI and IBIS we have no detailed positional information about the $^{44}$Ti emission but the constraint on the expansion velocity derived above, together with some nucleosynthesis considerations, can be interpreted in the context of all observational data currently available. Since $^{44}$Ti is synthesized in explosive Si-burning, a strong spatial correlation between $^{56}$Fe and $^{44}$Ti is expected. So a good starting point for an interpretation of our $^{44}$Ti result is no doubt an inventory of all the iron recorded so far in Cas A.\\
\indent From XMM-Newton data, \citet{Willingale:2003} performed an extensive spectral analysis of Cas A and determined a shocked ejecta mass of 2.2\,M$_\odot$ and a swept-up CSM mass of 7.9\,M$_\odot$. The abundances they obtained from spectral modelling indicate a total X-ray emitting iron mass of 0.058\,M$_\odot$, which is a sizeable fraction of the canonical 0.1\,M$_\odot$ of iron supposedly ejected in core-collapse events. Accounting for pre-existing iron in the progenitor reduces the value to about 0.04\,M$_\odot$ for the initial solar metallicity, which still constitutes a substantial amount. If we consider that the reverse-shocked ejecta may contain some fraction of cold material \citep[as suggested by][]{Hwang:2003}, presently undetectable in X-rays because of an earlier reverse-shock crossing and subsequent cooling or because of a lower density, the work of \citet{Willingale:2003} seems to indicate that a fair amount of the iron produced by Cas A currently lies between the forward and reverse shocks, with rms radial velocities of about 1700-1800 km\,s$^{-1}$.\\
\indent The total iron yield of the Cas A event remains however uncertain. The low brightness of the event recorded \citep[or not, see][]{Stephenson:2005} by Flamsteed in the 1680s suggests a low $^{56}$Ni yield, but the visual extinction towards Cas A might range from 4 to 8 magnitudes. Taking this into account, \citet{Young:2006} estimated that the maximum $^{56}$Ni mass that could have been produced in the event is 0.2\,M$_\odot$. The recent identification of Cas A as the remnant of a SNIIb explosion has allowed us to tighten the constraints on the total iron yield; it very likely lies between 0.07 and 0.15\,M$_\odot$, as indicated by the canonical SNIIb SN1993J \citep[see][and references therein]{Krause:2008}. So in the most optimistic scenario, Cas A harbours about 0.09-0.011\,M$_\odot$ in addition to the 0.04-0.06\,M$_\odot$ estimated by \citet{Willingale:2003}.\\
\indent From nucleosynthesis considerations, a substantial fraction of these 0.09-0.11\,M$_\odot$ of iron can be expected to lie inside the reverse shock. Indeed, in a spherical supernova explosion $^{56}$Fe is thought to form the deepest ejecta. According to the analytic hydrodynamic model used by \citet{Laming:2003} and \citet{Hwang:2003}, 0.2 M$_\odot$ of material are still freely expanding, so the most of iron could still be there, possibly accompanied by $^{44}$Ti. The free expansion velocity at the reverse-shock position was estimated to around 5000 km\,s$^{-1}$ by \citet{Morse:2004} from nearly undecelerated optical clumps. Therefore, the expansion velocities of the unshocked $^{56}$Fe and $^{44}$Ti ejecta should range between 0 and 5000 km\,s$^{-1}$.\\
\indent An alternative has been put forward in the particular context of $^{44}$Ti production. \citet{Maeda:2003} showed that a bipolar supernova explosion caused by a pair of opposite hydrodynamic jets fed by accretion brings the material on the jet axis to higher temperatures than that reached in spherical models, and hence causes a more rapid expansion and hence a stronger alpha-rich freeze-out for that material. This results primarily in an augmented $^{44}$Ti yield, but the hydrodynamics associated with that scenario also leads to an inversion of the velocities along the jet where $^{44}$Ti and $^{56}$Ni have the highest velocities around 20000 km\,s$^{-1}$, followed by $^{28}$Si and $^{16}$O. The simulation of \citet{Maeda:2003} has been invoked to explain the high $^{44}$Ti mass and $^{44}$Ti/$^{56}$Ni ratio inferred for Cas A \citep{Prantzos:2004} and also provides an explanation for the outward increasing iron abundance suggested by \citet{Laming:2006} and the outward increasing sulfur-to-oxygen ratio reported by \citet{van-den-Bergh:1971a} and \citet{Fesen:2001} for the optical FMKs of the jet. Moreover, \citet{Laming:2003} showed from a comparison of Chandra spectra of individual knots with self-similar hydrodynamical simulations that more energy was injected in the direction of the north-east jet. In the specific case of Cas A, the jet lies within $\pm$20$^{\circ}$ of the plane of the sky \citep{Fesen:2001}, so the 20000 km\,s$^{-1}$ of the $^{44}$Ti ejecta in this jet-induced nucleosynthesis scenario translates into maximum velocities along the line of sight of about $\pm$ 6400 km\,s$^{-1}$.\\
\indent The $^{44}$Ti ejecta can therefore be present in three distinct locations: the freely-expanding ejecta, the reverse-shocked ejecta and the jet. All three sites are associated with expansion velocities that are far higher than our $\sim$ 500\,km\,s$^{-1}$ lower limit and so the only conclusion we can draw from our INTEGRAL/SPI observations is that the $^{44}$Ti does not lie strictly at the mass cut (the asymptotic velocity of which is theoretically zero), contrary to what is obtained in spherically symmetric models.

% Conclusion
\section{Conclusion}
\label{conclusion}

We have searched for the three $^{44}$Ti decay lines in Cassiopeia A with the high-resolution spectrometer SPI on board INTEGRAL. Due to the very high instrumental background noise with strong time variability, it was not possible to extract the two lines at 67.9 and 78.4\,keV. For the high-energy line at 1157.0\,keV, no significant signal is seen in the 1140-1170\,keV band, suggesting that the 1157.0\,keV line signal from Cassiopeia A is broadened by the Doppler effect. From our spectrum, we derive a $\sim$ 500\,km\,s$^{-1}$ lower limit at 2$\sigma$ on the expansion velocity of the $^{44}$Ti ejecta.\\
\indent When interpreted in the context of all observational data currently available, especially in optical and X-rays, this result does not allow us to constrain the nucleosynthesis site of $^{44}$Ti or the explosion process since the velocities involved throughout the remnant are all far above our lower limit. Further progress on this issue may follow from renewed efforts to understand and model the background noise of the SPI instrument at low energies. In the medium term, interesting results could also arise from the search for the $^{44}$Sc K$\alpha$ fluorescence emission that follows the electron capture of $^{44}$Ti. Investigations with Chandra as reported by \citet{Theiling:2006} have been fruitless, but line detection for specific regions may be achieved through extended observations.

\begin{acknowledgement}
The SPI project has been completed under the responsibility and leadership of CNES. We are grateful to ASI, CEA, CNES, DLR, ESA, INTA, NASA and OSTC for their support.
\end{acknowledgement}

\bibliographystyle{aa}
\bibliography{/Users/pierrickmartin/Documents/MyPapers/biblio/CassiopeeA,/Users/pierrickmartin/Documents/MyPapers/biblio/44Ti,/Users/pierrickmartin/Documents/MyPapers/biblio/SNmodels,/Users/pierrickmartin/Documents/MyPapers/biblio/Nucleosynthesis,/Users/pierrickmartin/Documents/MyPapers/biblio/SNobservations,/Users/pierrickmartin/Documents/MyPapers/biblio/SPI,/Users/pierrickmartin/Documents/MyPapers/biblio/26Al&60Fe,/Users/pierrickmartin/Documents/MyPapers/biblio/SNRobservation}

\end{document}